\documentclass{PoS}
\def\Journal#1#2#3#4{{#1} {\bf #2}, #3 (#4)}

\title{Searching for optical and VHE counterparts of fast radio bursts with MAGIC}

\ShortTitle{MAGIC observations of FRBs}

\author{\speaker{J. Hoang}$^{1}$, M. Will$^{2}$, S. Inoue$^{3}$, J. A. Barrio$^{1}$, J. Cortina$^{4}$, M. L\'{o}pez$^{1}$, B. Marcote$^{5}$, L. A. Tejedor$^{1}$, on behalf of the MAGIC Collaboration\thanks{ https://magic.mpp.mpg.de, \newline \hspace*{7mm}For collaboration list see PoS(ICRC2019)1177}\\
       $^{1}$Universidad Complutense de Madrid, Grupo de Altas Energ\'{i}as (GAE)\\
       $^{2}$Max-Planck-Institut f{\"u}r Physik, M\"unchen\\
       $^{3}$RIKEN\\
       $^{4}$Centro de Investigaciones Energ\'{e}ticas, Medioambientales y Tecnol\'{o}gicas (CIEMAT)\\
       $^{5}$Joint Institute for VLBI ERIC (JIVE)\\
       E-mail: \email{kimhoang@ucm.es}}

\abstract{Fast radio bursts (FRBs) are an enigmatic class of extragalactic transients emitting Jy-level radio bursts in the GHz band, lasting for only a few ms. So far, some objects are known to repeat while several others are not, likely indicating multiple origins. There are many theoretical models, some predict prompt VHE or optical emission correlated with FRBs while others imply VHE afterglows hours after the FRB. To test these predictions and unravel the nature of FRB progenitors, the stereoscopic Imaging Atmospheric Cherenkov Telescopes (IACTs) system MAGIC has been participating in FRB observation campaigns since 2016. As IACTs are sensitive to Cherenkov photons in the UV/blue region of the electromagnetic spectrum and use photo-detectors with time response faster than a ms, MAGIC is also able to perform simultaneous optical observations through a dedicated system installed in the central PMT of its camera. The main challenge faced by MAGIC in searching for optical counterpart of FRBs is the presence of irreducible background optical events due to terrestrial sources. We present new results from MAGIC observations of the first repeating FRB 121102 during several MWL observation campaigns. The recently improved instrument and refined strategy to search for counterparts of FRBs in the VHE and optical bands will also be presented.
}

\FullConference{36th International Cosmic Ray Conference -ICRC2019-\\
		July 24th - August 1st, 2019\\
		Madison, WI, U.S.A.}

\begin{document}

\section{Introduction}
Fast Radio Bursts (FRBs) are bright but brief radio transients, typically detected at GHz frequencies, with duration of milliseconds and a peak flux up to $\sim$100 Jy (for reviews, see~\cite{KATZ,KEANE}). They exhibit clear dispersion effects, i.\,e., the arrival time of radio waves from FRB shows a characteristic, frequency-dependent delay, caused by propagation through intervening ionized matter. Their observed dispersion measures (DMs), reflecting the column density of free electrons along their path, are significantly larger than that expected for the Galactic interstellar medium. This strongly suggests their extragalactic origin, with their large DMs predominantly due to the intergalactic medium. After the first FRB was detected serendipitously with the Parkes telescope~\cite{LORIMER}, dedicated searches have uncovered more than 50~FRB sources to date\footnote{http://frbcat.org}. The estimated FRB event rate over the whole sky is on the order of few thousands per day above a fluence of few Jy ms at $\sim$1 GHz~\cite{KEANE}.

The nature of FRBs remains uncertain. Most single-dish radio telescopes used to observe FRBs have insufficient localization capabilities to achieve unambiguous association with counterparts at other wavelengths. As of July 2019, among the 3 FRBs that have been spatially localized by the use of long baseline interferometry ~\cite{CHATTERJEE, BANNISTER, RAVI}, FRB~121102 appears to be the only one showing clear repetition, disproving cataclysmic models of the source progenitors and suggesting multiple classes of origin. Its repeating nature simplifies the localization process, facilitates follow-up observations as well as MWL searches~\cite{CHATTERJEE,MARCOTE}. The source has been identified with a star-forming region in a low-metallicity dwarf galaxy at redshift $z\simeq 0.19$~\cite{BASSA,TENDULKAR}. Subsequent radio observations revealed large and variable Faraday rotation of linearly polarized burst emission, indicating an extreme, dynamic magneto-ionic environment~\cite{MICHILLI} that may point to a neutron star origin~\cite{HESSELS}. Further observations at other wavelengths are necessary to clarify the nature of FRB~121102. In particular, some theoretical models propose a young pulsar or magnetar as its progenitor predict optical~\cite{LYUTIKOV} or very-high-energy (VHE; above several tens of GeV) $\gamma$-ray emission~\cite{LYUBARSKY,MURASE} correlated in time with radio bursts. This can be tested by observing FRB~121102 with the MAGIC telescopes simultaneously with radio telescopes.

\section{Instrumentation}
MAGIC is a system of two Imaging Atmospheric Cherenkov Telescopes (IACT) with tessellated reflectors of 17-meter diameter located at the Roque de los Muchachos Observatory in La Palma, Canary Islands, Spain. The first telescope (MAGIC-I) was built in 2004, and operated for five years in mono mode. The second MAGIC telescope (MAGIC-II), built at a distance of 85\,m from the first, started taking data in July 2009. During 2011 and 2012, MAGIC underwent a major upgrade in two stages. In 2011, the readout electronics of the two telescopes was upgraded. In 2012, the camera of the MAGIC-I was replaced with a uniformly pixelized camera similar to the one mounted on MAGIC-II. Today, the two telescopes operate together in stereo mode, with the two cameras having fast imaging capability and a field of view (FoV) of 3.4~degrees, sensitive to photons in the U-band (365\,nm wavelength). Astrophysical VHE $\gamma$-rays interact with the atmosphere and trigger showers of secondary charged particles, which in turn emit Cherenkov radiation in the optical to UV range. By stereoscopically imaging this radiation, MAGIC can indirectly detect $\gamma$-rays in the energy range between 50\,GeV and 50\,TeV~\cite{ALEKSIC}.

Furthermore, the modified readout electronics of the photomultiplier tube at the camera center of MAGIC-II, known as the Central Pixel System (CPS), enables MAGIC to simultaneously read out optical signals at a rate of 10\,kHz. This modification makes it suitable for high time resolution photometry in the visible range of the electromagnetic spectrum. The CPS can detect periodic pulsations from the Crab pulsar in less than 10~seconds using the phase-folding technique. It is also able to detect isolated 10-ms optical flashes as faint as 13.4\,mag ($\sim$8\,mJy) with a sensitivity peaking at 350\,nm~\cite{HASSAN_ICRC}. On the other hand, the CPS is not sensitive to persistent optical emission, as its electronics is AC-coupled to the kHz frequencies and removes the persistent DC component of the signal.

\section{Observations}
In order to search for optical and VHE counterparts of FRB~121102, we observed the source with the MAGIC telescopes, some observation hours were carried out simultaneously with scheduled runs by the Arecibo radio telescope. Observations were conducted during 14~nights between September 2016 and September 2017, totaling 23\,h of good quality data after selection cuts, of which 8.9\,h were concurrent with Arecibo. MAGIC observations were carried out in the so-called ''ON mode''\footnote{The standard observation mode for MAGIC is the so-called ''wobble mode'', in which the telescopes point to a nearby direction, offset by 0.4\,deg to estimate the background for VHE $\gamma$-ray observations at the same time.}, i.\,e., with the source always located at the center of the FoV to allow optical observations using the CPS simultaneously with VHE observations~\cite{HASSAN_ICRC}. 

We conducted both a targeted search within 10\,ms windows around the Time of Arrival (ToA) of the FRB, and a blind search during the entire campaign. In addition, we also searched for persistent VHE $\gamma$-ray emission. When searching for brief, ms VHE $\gamma$-rays counterparts, the principal background is Cherenkov radiation due to isolated muons in air showers triggered by cosmic ray hadrons (mainly protons, with smaller amounts of helium and heavier nuclei). Although such background events dominate, $\gamma$-rays can be efficiently distinguished through differences in their Cherenkov images, and almost no background events are expected during intervals as short as ms. Thus, even a few events clustered within a few ms time window around an FRB would strongly indicate associated $\gamma$-rays emission from the FRB. The corresponding background was estimated using data collected during observations of other sources under similar conditions for the the night-sky background level and zenith angle.

For the optical band, we performed two kinds of searches for optical emission associated with FRB: a targeted search within 10\,ms time windows around the FRB ToAs, and a blind search during the entire observing campaign. For the targeted search, the main sources of background are optical flashes due to meteors traversing the Central Pixel FoV. Due to the low frequency of these meteors, transient optical pulses coincident with FRBs would be strong evidence of optical signal from the source. For an unbiased blind search for 10-ms optical pulses around the radio ToAs, we sequentially increased the total search window around each FRB ToA in equal logarithmic time steps (i.\,e., 10\,ms, then 100\,ms, 1\,s and so on). Unlike the VHE regime, a persistent search for optical emission with variability scale greater than one second was not possible since the AC-coupled Central Pixel removes the persistent DC component of the signal.

\section{Data analysis and Results}
Five FRBs from FRB~121102 were detected by Arecibo during this campaign when the MAGIC observing conditions were favorable, with atmospheric transmission at 96\% and zenith angles less than 46~degrees~\cite{MAGIC_FRB}. The ToA, DM, and peak radio flux for these 5~FRBs, and the corresponding MAGIC observing conditions, are listed in Table~\ref{tab:frb_ToAs}. ToAs at the MAGIC site have been corrected to infinite frequency and for the different expected topocentric times. These corrections are less than 10\,ms and vary among the 5~FRBs.

\begin{table}
\tiny
\caption{FRBs detected by Arecibo during MAGIC observation campaign, together with the observing conditions for MAGIC at the corresponding epochs. The reported aerosol transmission refers to the atmospheric optical depth relative to a standard dark night.}

\begin{center}
\scriptsize
\begin{tabular}{c c c c c|c c c}
\hline
 MJD (Arecibo) & DM & Duration & Peak brightness & Significance (Arecibo) & MJD (MAGIC) & Transmission & Zd  \\
 {[days]}   &  [pc cm$^{-3}$]  &  [ms]  & [Jy]  & [$\sigma$]   & [days]  &   & [deg] \\ 
 \hline
 57799.98317566 & 562 & 5.73 & 1.4 & 32.17 & 57799.98316670 & 0.96 & 33 \\ 
 57806.96425078 & 562 & 2.46 & 1.6 & 38.59 & 57806.96424183 & 0.96 & 33 \\ 
 57806.98472905 & 561 & 3.69 & 1.5 & 35.21 & 57806.98472011 & 0.96 & 40 \\ 
 57808.00278585 & 563 & 3.69 & 0.79 & 18.73 & 57808.00277693 & 0.96 & 46 \\ 
 57814.94698520 & 560 & 1.15 & 0.47 & 11.13 & 57814.94697625 & 0.96 & 35 \\
\end{tabular}
\end{center}
\label{tab:frb_ToAs}
\end{table}

\begin{figure}[ht]
\centering
\includegraphics[width=0.7\textwidth]{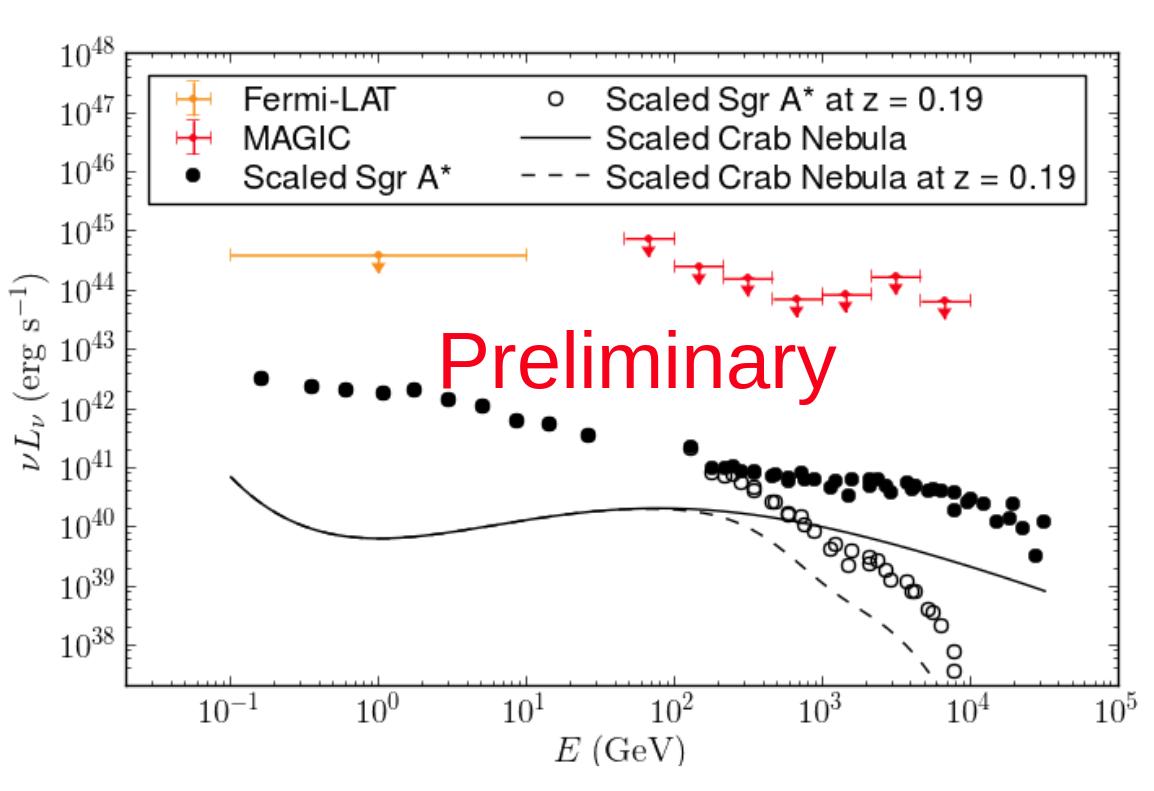}
\caption{Upper limits for the persistent emission of FRB~121102. See text for explanation}
\label{fig:UL_SED}
\end{figure}

VHE $\gamma$-rays data were analyzed using MARS, the standard MAGIC Reconstruction Software~\cite{MARS}. The recorded shower images were calibrated, cleaned, and parameterized according to standard Hillas parameters~\cite{HILLAS}. For the target search, with a time window of 10 ms centred around the radio burst ToAs and using special analysis cuts (optimized cuts in size, hadronness, and $\theta^{2}$ to maximize sensitivity over ms-duration observation times), no gamma-like events were found within any of these windows above 100\,GeV. No VHE events were detected within these ToAs, and the resulting upper limit for each FRB corresponds to 3.56~photons above 100\,GeV. We also performed the search for VHE bursts for the entire data set, but no hint of clusters with 3.56~photons within 10 ms window was found. No persistent VHE gamma-ray emission was detected from FRB~121102 (see also~\cite{BIRD}). Assuming a power-law spectrum with photon index $\Gamma$\,=\,2, integral flux upper limits for energy greater than 150, 400 and 1000 GeV for 23~hours are 2.50, 0.89, and 0.42\,$\times$\,10$^{-12}$\,cm$^{-2}$\,s$^{-1}$, respectively. The upper limits in luminosity for the persistent $\gamma$-ray emission of FRB 121102 from MAGIC (95\% confidence level, assuming an intrinsic power-law spectrum with $\Gamma$ = 2 and 30\% overall systematic uncertainty) up to 10\,TeV is also shown in Fig.~\ref{fig:UL_SED}. We also show limit from Fermi-LAT \cite{ZHANG}. To illustrate the effect of the intergalactic distance, the SEDs of the Crab Nebula \cite{MEYER} and Sgr $A^{\star}$ \cite{ABDO,AHARONIAN} have been scaled by factors of 4 $\times$ 10$^5$ and 2 $\times$ 10$^6$ to match the observed radio luminosity of the persistent radio source associated with FRB 121102. Additionally, both are shown with the effect of attenuation of the gamma rays due to interactions with the extragalactic background light \cite{DOMINGUEZ}, which is significant above $\sim$400 GeV at the estimated redshift z $\sim$0.19 of FRB 121102. 

\begin{figure}
\begin{center}
\includegraphics[width=0.8\textwidth]{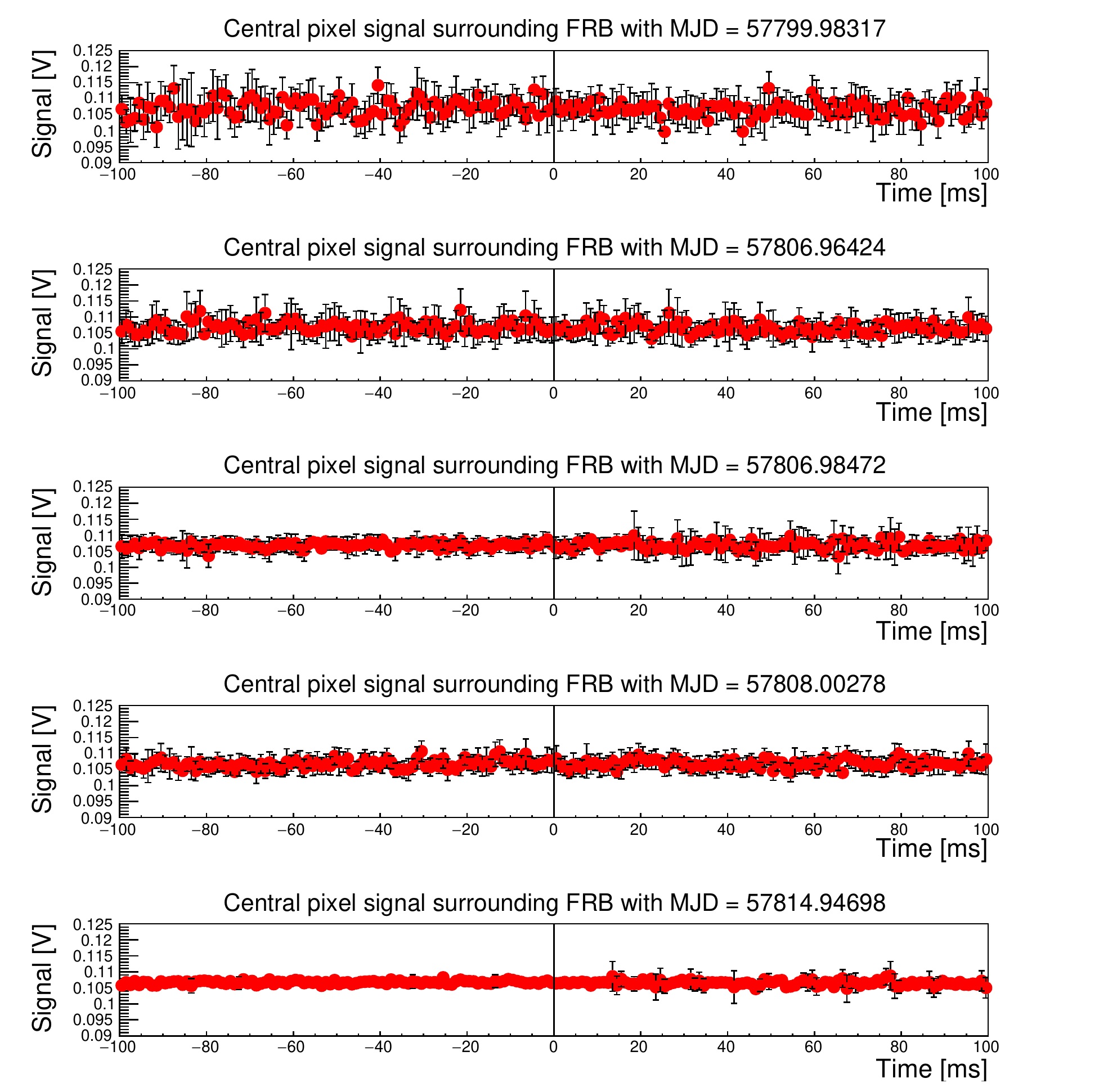}
\caption{Optical light curves around the 5 ToAs detected by the Arecibo telescope during observations simultaneous with MAGIC. The vertical axis is proportional to the photon flux in the U-band. No isolated optical pulse on ms timescales is detected simultaneously with any of the 5 bursts. The noise level varies with the sky brightness. Figure taken from~\cite{MAGIC_FRB}.}
\label{fig:Optical_TOA}
\end{center}
\end{figure}

As shown in Fig.~\ref{fig:Optical_TOA}, no significant optical emission was detected simultaneously with any of the five FRBs (see also \cite{HARDY}). An isolated optical pulse with brightness $\sim$29\,mJy\,$\sim$13\,mag was clearly observed $\sim$\,4\,s before the ToA of one FRB. No other optical pulses were detected near the ToAs of the remaining four FRBs. Considering the frequency of background events that are not reliably rejected in our analysis, an optical pulse of this brightness is statistically consistent with the expected background (2.2$\sigma$, post-trial). More details of these observations, the implications of the obtained upper limits for constraining FRB models and the prospects for future observations can be found in ~\cite{MAGIC_FRB}.

\section{Improved instrumentation and refined strategy}
A second Central Pixel for the MAGIC-I telescope is currently under development, allowing optical observation to be carried out in stereo mode. Stereo observation allows MAGIC to efficiently reject fast optical pulses resulted from terrestrial cause such as car flashes, interferences from several laser calibration systems from MAGIC itself and other observatories present in La Palma, satellites, and space debris. Additionally, assuming that scintillation noise is not correlated at 8\, m given MAGIC's large mirror size and the dominant noise will thus be dominated only by the NSB, the two-fold increase in the number of data points will improve the sensitivity by a factor of $\sqrt{2}\,\sim\,1.4$, which will enhance the significance in case of optical counterpart detection. Finally, to effectively discriminate irreducible terrestrial optical backgrounds from meteors traversing through MAGIC's FoV, future searches for FRB optical counterparts may be carried out simultaneously with the VERITAS telescopes (which, similar to MAGIC, also posses high-time-resolution optical photometry capability \cite{BENBOW}) to effectively discriminate irreducible terrestrial optical backgrounds from meteors. 

Recently during its pre-commissioning phase, the CHIME telescope discovered the second repeating FRB (FRB~180814), which is closer to Earth by a factor of approximately 2 as compared to FRB~121102, $z\,\simeq$\,0.11 and 0.19, respectively~\cite{CHIME}. CHIME's sensitivity suggests that a substantial population of repeating FRBs will be detected in the near future, and this important discovery motivates MAGIC's continued effort to search for optical and VHE counterpart of FRBs. Even though the estimated rate for FRB is on the order of 10$^4$~events per sky per day, some FRBs have been observed for more than 10$^3$ hours without any re-detection~\cite{SHANNON}. The repeaters, on the other hand, seem to repeat on the timescale between hours and days during their episodic outbursts. To optimize the detection efficiency within MAGIC's observation time constraints, we follow two strategies to search for FRB counterparts with MAGIC. To detect non-delayed afterglow emission similar to the cataclysmic case of Gamma-Ray Bursts, MAGIC will perform automatic follow-up observations of newly discovered non-repeating FRBs, utilizing its capability to point to any position of the sky within less than 30 seconds. The pointing will be made as soon as possible after a detection of new FRBs, triggered via an automatically parsed and filtered alert based on the Virtual Observatory Events (VOEvents) that is currently in development. Observation on other repeating FRBs, on the other hand, will be activated via Target of Opportunity mechanism whenever the sources enter their outburst period. The alert will be received through human-readable format such as ATel or custom communication channel per Memorandum of Understanding (MoU) agreement.     

\section{Acknowledgements}

\noindent
\href{https://magic.mpp.mpg.de/acknowledgments&#95;ICRC2019/}{https://magic.mpp.mpg.de/acknowledgments\_ICRC2019/}

\noindent
In addition, the authors would like to thank Tarek Hassan, Jason Hessel, Daniele Michilli, and our colleagues from the Radio communities, for the very fruitful exchanges we had on FRB 121102.

\end{document}